\documentclass[usenatbib,twocolumn]{aastex61}
\usepackage{hyperref}
\pdfoutput=1

\newcommand\starcountall{213}
\newcommand\starcountspectra{365}
\newcommand\starcountbinary{16}
\newcommand\starcountrotate{57}
\newcommand\starcountsample{140}

\received{2018 March 1}
\accepted{2018 April 1}
\submitjournal{PASP}

\shorttitle{Spectroscopy of dwarf stars around the north celestial pole}
\shortauthors{Mikolaitis et al.}

\begin{document}

\title{Spectroscopy of dwarf stars around the north celestial pole \footnote{Based on observations collected with the 1.65 m telescope and VUES spectrograph at the Mol\.{e}tai Astronomical Observatory of Institute of Theoretical Physics and Astronomy, Vilnius University, for the SPFOT survey.}}

\correspondingauthor{\v{S}ar\={u}nas Mikolaitis}
\email{sarunas.mikolaitis@tfai.vu.lt}

\author[0000-0002-1419-0517]{\v{S}ar\={u}nas Mikolaitis}
\affil{Institute of Theoretical Physics and Astronomy, Vilnius University, Saul\.{e}tekio av. 3, 10257 Vilnius, Lithuania}

\author{Gra\v{z}ina Tautvai\v{s}ien\.{e}}
\affiliation{Institute of Theoretical Physics and Astronomy, Vilnius University, Saul\.{e}tekio av. 3, 10257 Vilnius, Lithuania}

\author{Arnas Drazdauskas}
\affiliation{Institute of Theoretical Physics and Astronomy, Vilnius University, Saul\.{e}tekio av. 3, 10257 Vilnius, Lithuania}

\author{Renata Minkevi\v{c}i\={u}t\.{e}}
\affiliation{Institute of Theoretical Physics and Astronomy, Vilnius University, Saul\.{e}tekio av. 3, 10257 Vilnius, Lithuania}

\author{Lukas Klebonas}
\affiliation{Institute of Theoretical Physics and Astronomy, Vilnius University, Saul\.{e}tekio av. 3, 10257 Vilnius, Lithuania}
\affiliation{Mathematisch-Naturwissenschaftliche Fakult\"at, Universit\"at Bonn, Wegelerstra{\ss}e 10, 53115 Bonn, Germany}

\author{Vilius Bagdonas}
\affiliation{Institute of Theoretical Physics and Astronomy, Vilnius University, Saul\.{e}tekio av. 3, 10257 Vilnius, Lithuania}

\author{Erika Pak\v{s}ien\.{e}}
\affiliation{Institute of Theoretical Physics and Astronomy, Vilnius University, Saul\.{e}tekio av. 3, 10257 Vilnius, Lithuania}

\author{Rimvydas Janulis}
\affiliation{Institute of Theoretical Physics and Astronomy, Vilnius University, Saul\.{e}tekio av. 3, 10257 Vilnius, Lithuania}



\begin{abstract}

New space missions (e.g. NASA-TESS and ESA-PLATO) will perform an in-depth analysis of bright stars in large fields of the celestial sphere searching for extraterrestrial planets and investigating their host-stars. Asteroseismic observations will search for exoplanet-hosting stars with solar-like oscillations. In order to achieve all the goals, a full characterization of the stellar objects is important. However, accurate atmospheric parameters are available for less than 30\% of bright dwarf stars of the Solar neighborhood. In this study we observed high-resolution (R=60000) spectra for all bright ($V<8$~mag) and cooler than F5 spectral class dwarf stars in the northern-most field of the celestial sphere with radius of 20 degrees from the $\alpha$(2000)~=~161.03$^\circ$ and $\delta$(2000)~=~86.60$^\circ$ that is a centre of one of the preliminary ESO-PLATO fields. 
Spectroscopic atmospheric parameters were determined for \starcountsample~slowly rotating stars, for 73\% of them for the first time. 
The majority (83\%) of the investigated stars are in the TESS object lists and all of them are in the preliminary PLATO field. Our results have no systematic differences when compared  with other recent studies. We have 119 stars in common with the Geneva-Copenhagen Survey, where stellar parameters were determined photometrically, and find a $14\pm125$~K difference in effective temperatures, $0.01\pm0.16$ in log\,$g$, and $-0.02\pm0.09$~dex in metallicities. Comparing our results for 39 stars with previous high-resolution spectral determinations, we find only a $7\pm73$~K difference in effective temperatures, $0.02\pm0.09$ in log\,$g$, and $-0.02\pm0.09$~dex in metallicities. 
We also determined basic kinematic and orbital parameters for this sample of stars.
From the kinematical point of view, almost all our stars belong to the thin disk substructure of the Milky~Way. The derived galactocentric metallicity gradient is $-0.066\pm0.024$~dex~kpc$^{-1}$ (2.5\,$\sigma$ significance) and the vertical metallicity gradient is
$-0.102\pm0.099$~dex~kpc$^{-1}$ (1$\sigma$ significance) that comply with  the latest inside-out thin disk formation models, including those with stellar migration taken into account.

\end{abstract}




\section{Introduction}
\label{sec:intro}

Stars that are in a close proximity to our Solar system are the ones that can be studied in detail using good quality of astrometric, photometric and spectroscopic measurements.  In the era of new planetary search space missions (e.g. NASA-TESS or ESA-PLATO) bright stars in the Solar neighborhood become even more important to investigate. If a planetary system is found around a bright nearby star it gets a large attention from many astrophysical perspectives. A full asteroseismic characterisation of a star is possible if its main atmospheric parameters are known. 

It could be thought that cool bright main sequence stars in the Solar neighborhood should have been carefully studied, however, only about 30\% of F5 and cooler dwarfs with magnitudes $V <$~8~mag have been studied spectroscopically. This is very unfortunate for asteroseismic and planetary studies. For example, stellar parameters and precise elemental abundances are key constraints in reducing degeneracy in the interior structure models and inspecting the mantle composition of exoplanets (\citealt{Weiss2013,Dorn2015,Dorn2017}). The precise and accurate atmospheric physical and chemical parameters of stars are required in order to produce trustworthy results. Therefore, ground-based support for space missions is very important.

Large spectroscopic surveys have a great input in studying properties of Solar environment and far beyond, however their observational strategies usually exclude most of the bright nearby stars. For example, surveys like Gaia-ESO \citep{Gilmore2012} or APOGEE \citep[]{Majewski2017} cover limited fractions of the sky and exclude bright nearby stars. Such surveys like GALAH (De Silva et al. 2015), 4MOST (de Jong et al. 2011), and WEAVE (Dalton et al. 2016) will cover huge fractions of the sky, however will be generally concentrating on stars fainter than $V$=12~mag. 

Large telescopes cannot pay much attention in surveying nearby bright stars, leaving the study of Solar neighborhood mainly to archival data (cf. \citealt{Adibekyan2012}, \citealt{Worley2012, Worley2016, DePascale2014}), however all these samples cover only a small fraction of the sky.
Such spectroscopic surveys of stars in the Solar neighborhood like \citet{Prieto2004} who studied in a high-resolution mode all stars up to $V$=6.5~mag around the Sun (nearest 15~pc) are very rare. 

The metllicity gradients of the disk are important tools in studying the Galactic disk formation. Usually, this task lies on large spectroscopic surveys (e.g. \citealt{Coskunoglu2012, Boeche2013, Cheng2012, Duong2018}). The Solar proximity stars are not well spread to provide meaningful vertical or radial gradients. However, kinematic and orbital computations show the statistical positions of these stars during their movement around the Galaxy. The orbital parameters such as the mean galactocentric distance show the original spatial distribution of the presently neighboring stars and can be studied in the context of metallicity gradients (e.g. \citealt{Anders2014, Boeche2013}).

Therefore, bright ($V <$ 8~mag) stars are perfect objects for small and intermediate size telescopes that have high-resolution spectrographs installed.
A high-resolution spectrograph (\citealt{Jurgenson2014,Jurgenson2016}) that is installed at the Mol\.{e}tai Astronomical Observatory of the Institute of Theoretical Physics and Astronomy, Vilnius University (ITPA VU) on the largest telescope in the northern Europe gives us a chance to provide an input in understanding of stars located in the Solar vicinity towards the northern celestial pole.

Hence, in 2016 we started a Spectroscopic and Photometric Survey of the Northern Sky (SPFOT) that aims to provide a detailed chemical composition from high-resolution spectra and photometric variability data for bright stars in the northern sky. This is the first phase of the spectroscopic part of the effort where we describe characteristics of F5 and cooler main sequence bright ($V<8$~mag) stars that are located towards the northern celestial pole in a sky-field with radius of 20 degrees. The majority (83\%) of the observed targets are included in the TESS catalogs and have TESS designated IDs. The field of our first observational phase partly coincide with the TESS continuous viewing zone and one of the possible PLATO fields (\citealt{Miglio2017}).

In Section~\ref{sec:targets}, we outline the information about the selected field and the strategy of target selection.  Section~\ref{sec:observations} describes observations 
and  data reduction procedures. In Section~\ref{sec:methods}, we describe our methods of analysis and possible uncertainties. In Section~\ref{sec:parameters}, we present the determined atmospheric and kinematic parameters of stars and compare them with determinations reported in the literature. Finally, Section~\ref{sec:summary} summarizes the results and foresees the future work.

  \begin{figure}[htb]
   \advance\leftskip 0cm
   \centering
    \includegraphics[scale=0.32]{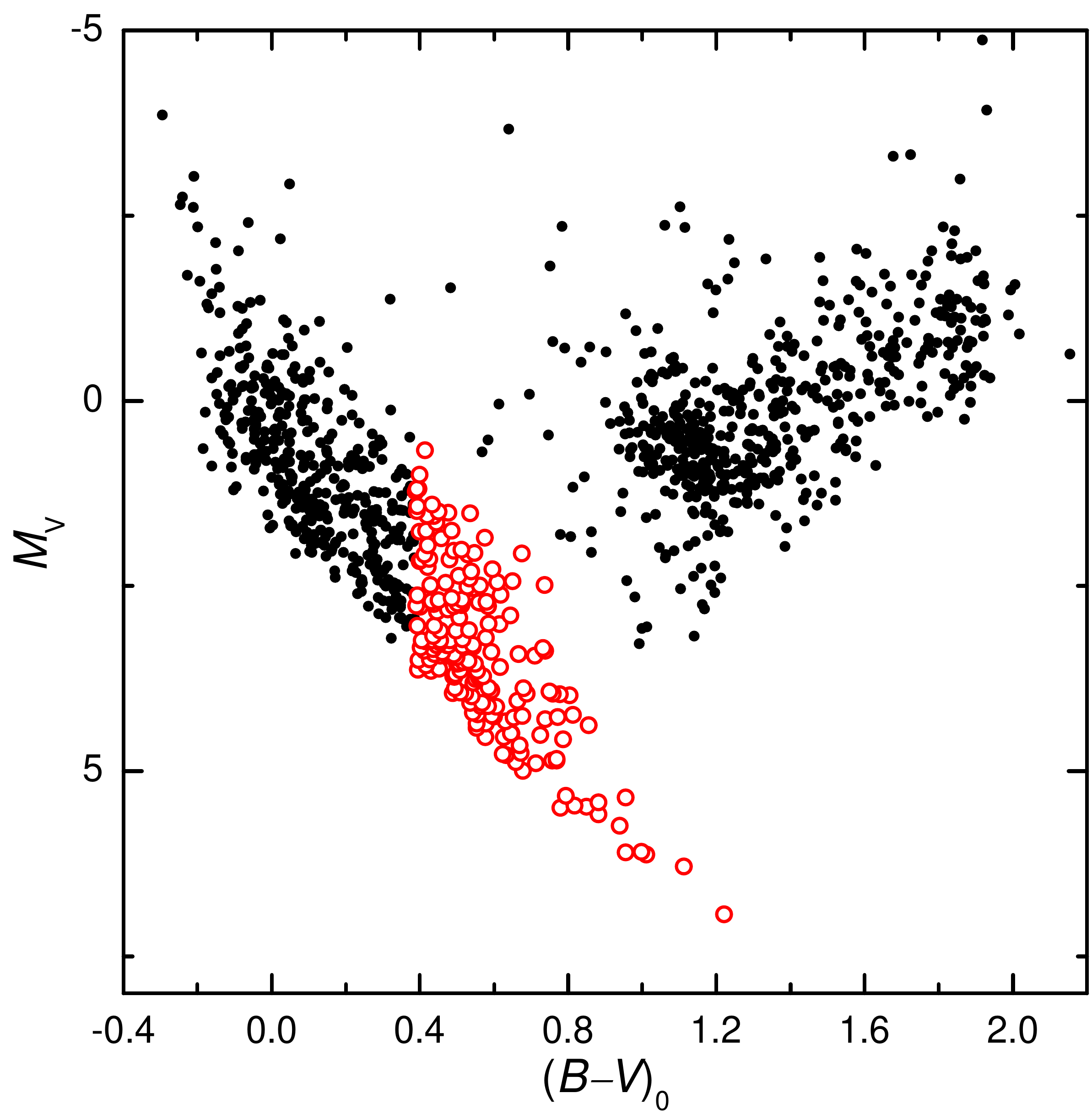}
  \caption{Color-magnitude diagram of stars in the investigated field. The FGK dwarfs observed in this programme are presented as red open circles.}
  \label{fig:cmd_targets}
  \end{figure} 

  \begin{figure}[htb]
   \advance\leftskip 0cm
   \centering
   \includegraphics[scale=0.36]{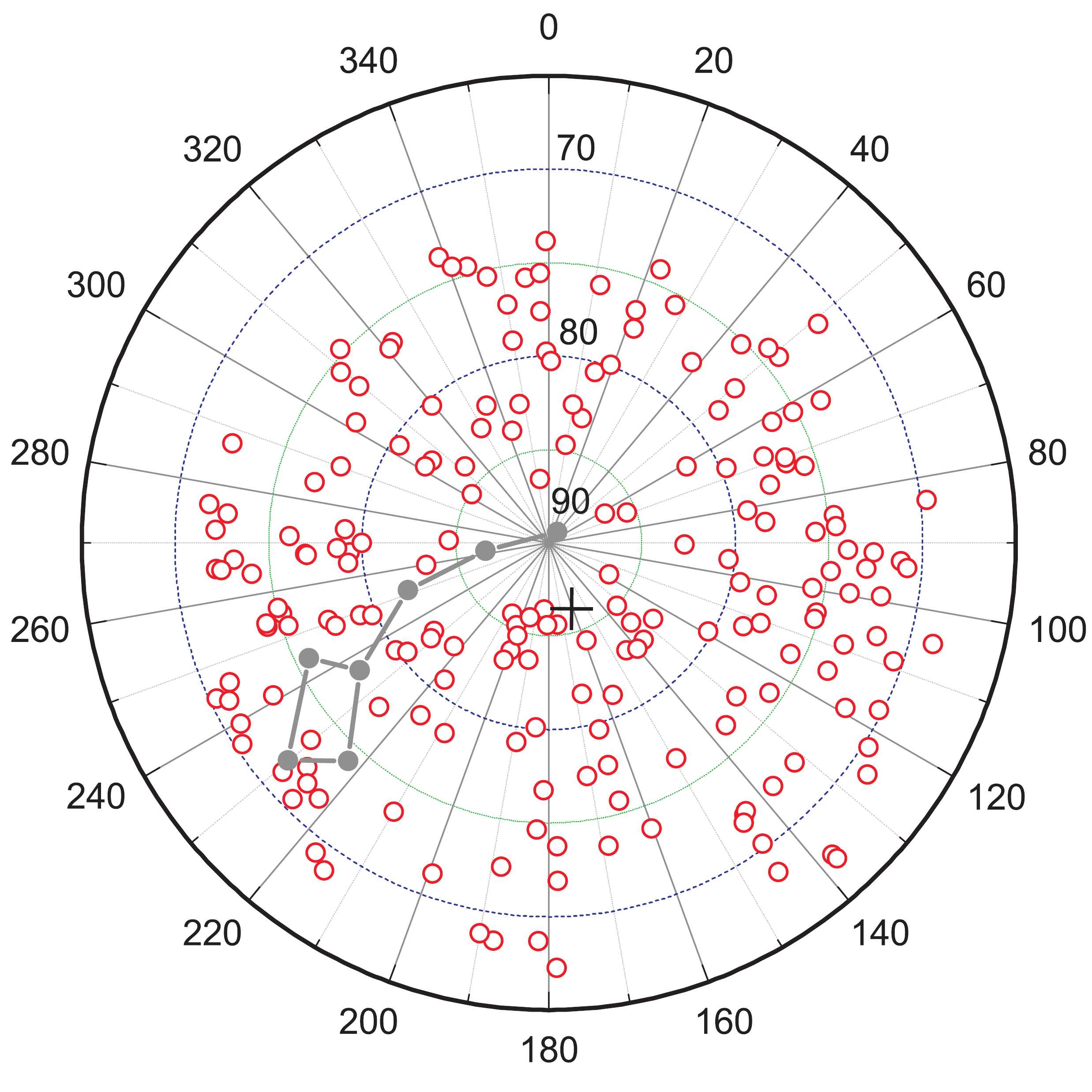}
  \caption{Postitions (RA and DEC in degrees) of the programme stars (red open circles). The centre of the field is shown as the black cross. }
  \label{fig:step02_map}
  \end{figure}   

  \begin{figure*}[htb]
   \advance\leftskip 0cm
   \centering
   \includegraphics[scale=0.62]{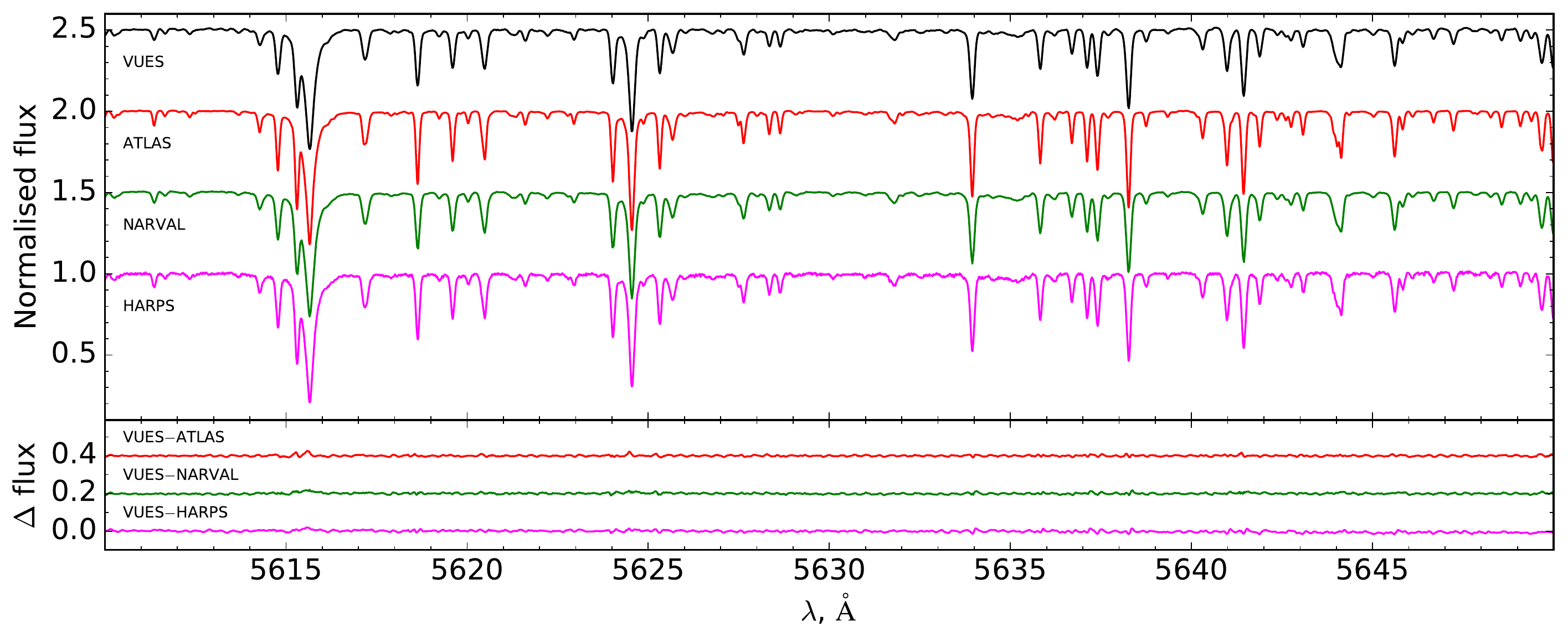}
  \caption{Upper panel: examples of the Solar spectra from the ATLAS (\citealt{Wallace2011}), HARPS (\citealt{Mayor2003}), NARVAL (\citealt{Auriere2003}), and VUES (\citealt{Jurgenson2016}) spectrographs (original resolutions). Lower panel: the flux differences between VUES and other spectra (the comparison spectra were downgraded to match the resolving power of VUES).}
  \label{fig:sollar_spectra}
  \end{figure*}   

\section{Target selection}
\label{sec:targets}

An initial list of 1163 stars was compiled from the TYCHO-2 catalog \citep{Hog2000} as follows. 
We searched the catalogue considering two main criteria. 
Firstly, we approximated the shape of the field to a disk with the radius of 20~degrees around the centre of the field.
Northern celestial pole is very close to the centre of one of the possible PLATO fields (\citealt{Rauer2014, Rauer2016})
thus we centered our observations accordingly (20~degrees around $\alpha$(2000)~=~161.03552$^\circ$ and $\delta$(2000)~=~86.60225$^\circ$). Secondly, we limited a number of stars by setting the visual magnitude $V<8$~mag, thus adjusting our research to the strategies of space missions and the best performance of the telescope.

Then we constructed a colour--magnitude diagram (CMD). For this purpose, we computed ${\rm (}B-V{\rm )}_{0}$ and $M_{V}$. 
Due to the large size of our field, it was difficult to investigate the extinction law abnormalities. Thus, according to the mean extinction law we adopted the $R=3.1$ value as an average for the whole field (e.g. \citealt{Fitzpatrick1999}).
The $E_{B-V}$ values were calculated using the model of large-scale visual interstellar extinction by \citet{Hakkila1997}. The parallaxes required for the stellar distance calculations were gathered from the HIPPARCOS Catalog (\citealt{Perryman1997}).
We have selected to observe 6500~K and cooler main sequence and sub-giant stars. It complies with the observational strategies of TESS and PLATO. The majority of the stars in the Solar neighborhood are more metal rich than $-0.5$~dex (c.f. \citealt{Prieto2004}). Using ${\rm [Fe/H]}=-0.5$ inside the color index and temperature relations by \citep{Alonso1996} we found that $T_{\rm eff}$~$\approx$~6500~K corresponds to  ${\rm (}B-V{\rm )}\approx0.39$.
Therefore, we have constructed a color-magnitude diagram for all selected 1163 stars in the field (see Figure~\ref{fig:cmd_targets}).
We chose only dwarf and subgiant stars with ($B-V$)~$>$~0.39 (red circles in Figure~\ref{fig:cmd_targets}) as our targets. The target list consists of \starcountall~objects, and we have observed all of them. Positions of the observed stars in a celestial sphere are shown in Figure~\ref{fig:step02_map}. 

\begin{table*}
\caption{Sample table of a summary of observations.}
\begin{tabular}{llccllc}

\hline\hline
Tycho2 ID	& TESS ID (TIC-5) & Date and UTC time & Exposure time (s) & RA* (h:m:s) &DEC* ($^\circ:':"$) & S/N\\
\hline
TYC 4634-2068-1 & 288183829 & 2016-06-23 01:35:49.755 & 1200 & 14:48:6.16 & +82:27:4.565 & 87 \\
TYC 4634-2068-1 & 288183829 & 2017-02-07 04:54:41.028 & 2400 & 14:49:5.702 & +82:25:23.522 & 87 \\
TYC 4615-1144-1 & 401497601 & 2016-08-25 03:14:50.853 & 2400 & 01:01:0.882 & +83:11:2.74 & 48 \\
TYC 4615-1144-1 & 401497601 & 2016-09-07 03:37:47.693 & 2400 & 01:00:57.999 & +83:11:0.075 & 48 \\
TYC 4615-1144-1 & 401497601 & 2016-10-18 00:09:01.201 & 2400 & 01:01:5.135 & +83:11:23.529 & 48 \\
\hline
 \label{tab:observations}
\end{tabular}

Notes. A full table is only available in an electronic form at the CDS.\\
Coordinates correspond to the star position at the time of the observation. \\
\end{table*}

\begin{table*}
\caption{Sample table of the kinematic properties of the observed stars.}
\begin{tabular}{llccllccccccccc}

\hline\hline
Tycho2 ID & $U_{\rm LRS}$ & $V_{\rm LRS}$ & $W_{\rm LRS}$ & $R_{\rm mean}$ & $z_{\rm max}$ & $e$ & $TD/D$ & $V_{\rm rad}$ & $\sigma_{V_{\rm rad}}$ & FWHM$_{V_{\rm rad}}$\\
\hline
TYC 4634-2068-1 & $-67.70$ & 23.45 & 19.42 & 9.14 & 0.37 & 0.24 & 0.11 & $-7.5$ & 3.3 & 7.7  \\
TYC 4615-1144-1 & 5.00 & 2.14 & $-14.92$   & 7.81 & 0.23 & 0.04 & 0.01 & $-7.4$ & 3.5 & 8.2 \\
\hline
 \label{tab:kinematics}
\end{tabular}

Notes. A full table is only available in an electronic form at the CDS.\\
\end{table*}

\section{Observations}
\label{sec:observations}

The programme stars were observed with the Vilnius University Echelle Spectrograph (VUES) designed and constructed at the Exoplanet Laboratory of the Yale University (\citealt{Jurgenson2014,Jurgenson2016}) and mounted on the f/12 1.65~meter Ritchey-Chretien telescope at the Mol\.{e}tai Astronomical Observatory of the Institute of Theoretical Physics and Astronomy, Vilnius University. The VUES is designed to observe spectra in the 4\,000~to~8\,800~\AA~wavelength range with three spectral resolution modes ($R$~=~30\,000, 45\,000, and 60\,000).
Figure~\ref{fig:sollar_spectra} shows an example of solar spectra obtained with the VUES ($R$~=~60\,000) and several other well known spectrographs (upper panel), the lower panel shows flux differences between the VUES and other spectrographs. 

During an observational period of 2016--2017 we obtained \starcountspectra~spectra of \starcountall~stars. For  stars that were observed several times, their spectra were combined in order to increase a signal-to-noise ratio. A sample of the observation log is provided in Table~\ref{tab:observations}.
Bias, flat field, and calibration lamp measurements were acquired every evening before observing stellar spectra. We used a quartz lamp for the flat fielding and the ThAr spectra for the wavelength calibration. The  data  were  reduced  and  calibrated  following  standard reduction procedures which included a subtraction of the bias frame, correction for flat field, extraction of orders, wavelength calibration, and a cosmic rays removal (\citealt{Jurgenson2016}).

\section{Method of Analysis}
\label{sec:methods}

\subsection{Radial velocity determination and identification of double-line binaries and fast-rotating stars}

\begin{figure}[!htb]
   \advance\leftskip 0cm
   \centering
   \includegraphics[scale=0.26]{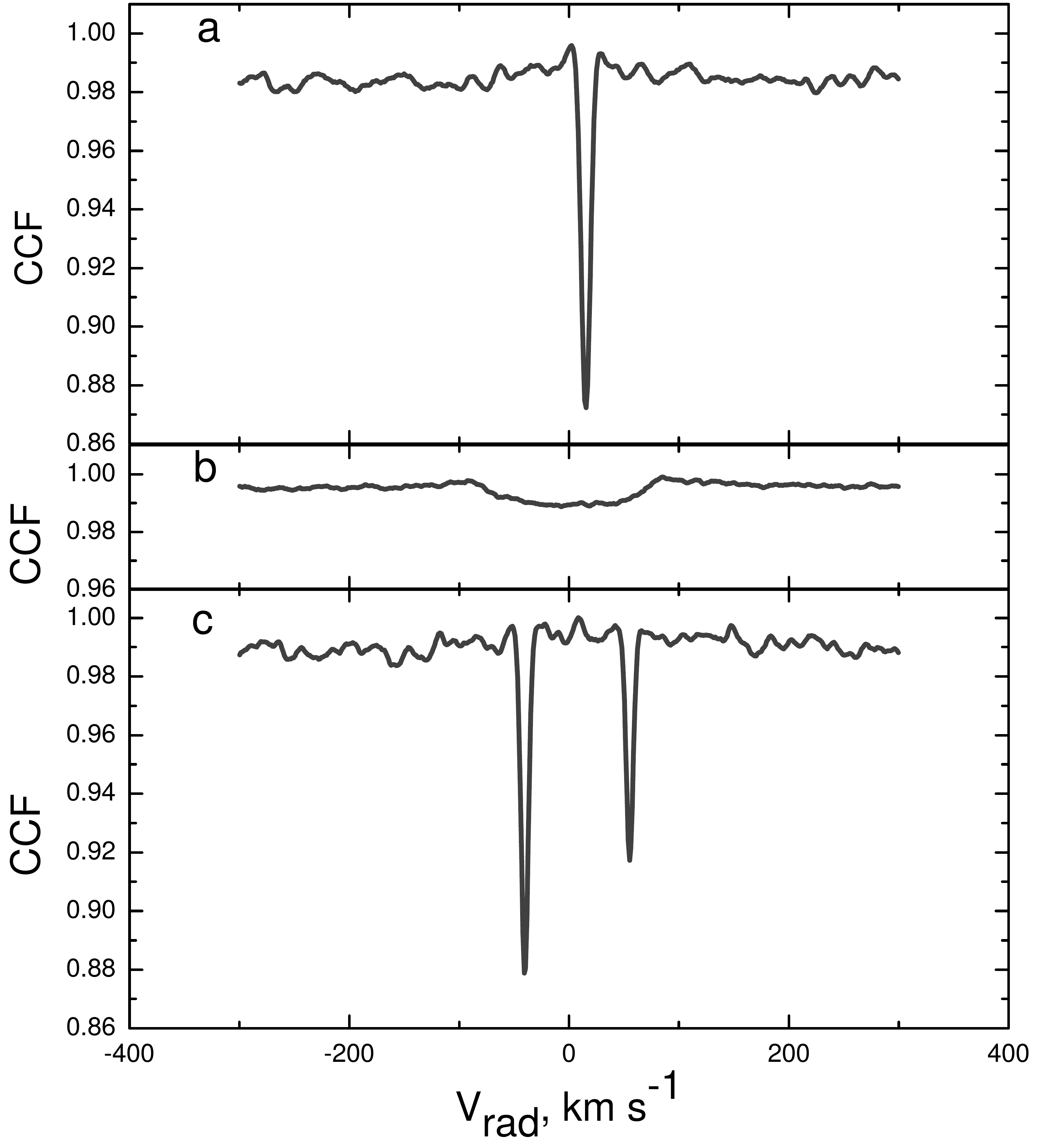}
  \caption{CCFs produced for calculating the radial velocities and detection of double-line binary stars: a) a typical slow-rotating star, b) a fast-rotating star, c) CCF of the double-line spectroscopic binary TYC~4602-552-1 (HIP~117712), showing two profiles.}
  \label{fig:ccf}
  \end{figure}

\begin{figure}[!htb]
   \advance\leftskip 0cm
   \centering
   \includegraphics[scale=0.22]{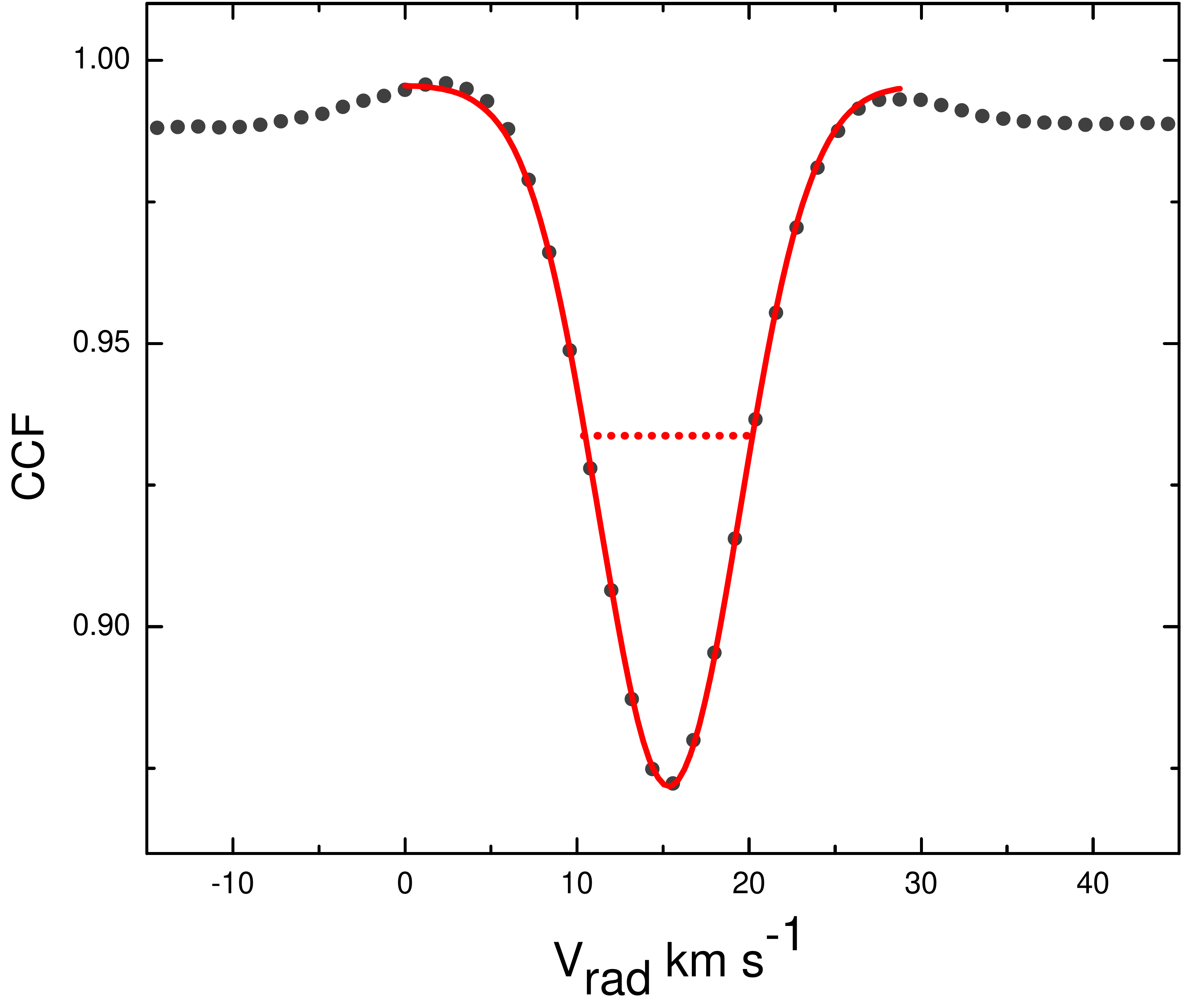}
  \caption{The same CCF function from Figure~\ref{fig:ccf}a with the fitted Gaussian function (red line), from which the radial velocities were determined. The FWHM of the Gaussian is shown as the red dotted line.}
  \label{fig:ccf_gauss_fit}
  \end{figure}

\begin{figure}[!htb]
   \advance\leftskip 0cm
      \centering
   \includegraphics[scale=0.26]{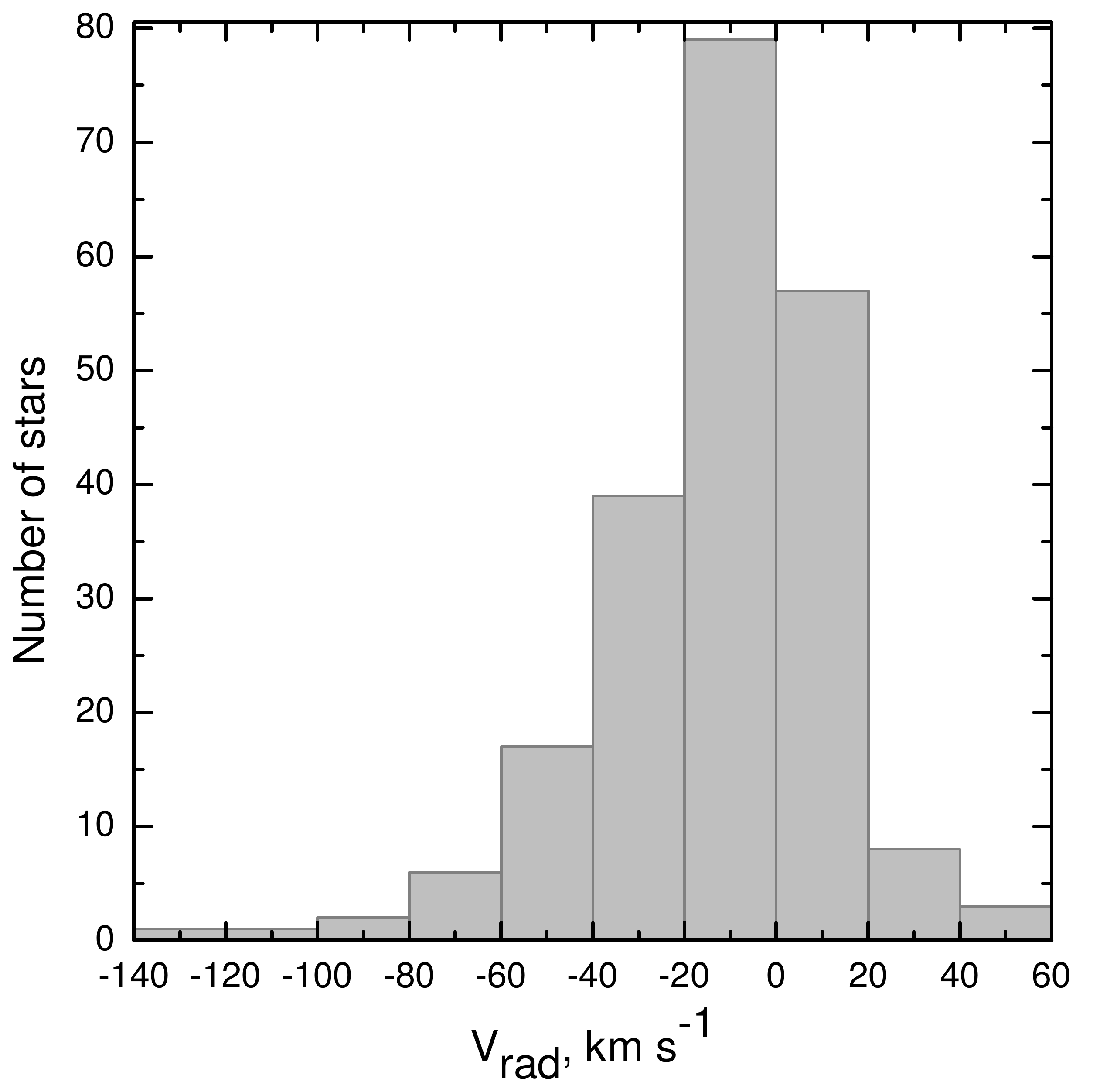}
  \caption{Histogram of the radial velocity values calculated for all targeted stars.}
  \label{fig:vrad_histo}
  \end{figure}

Since the target selection was done using only the photometric indices, a number of fast-rotating stars or spectroscopic double-line binaries were expected in the target list. Therefore, we performed a simple cross-correlation of spectra with a mask based on the atomic line list in order to detect double-line binaries and compute radial velocities. The line list for the mask was combined by the Gaia-ESO survey line list group (\citealt{Heiter2015}). It is composed of 1341 atomic transitions for 35 chemical elements. For each spectrum a cross-correlation function (CCF) ranging from $-300$ to +300 km\ s$^{-1}$ was calculated in radial velocity steps of $\Delta$V$_{rad}$=1.2 km\ s$^{-1}$. 
Figure~\ref{fig:ccf} shows the CCF corresponding to the fit for three spectra: a slow rotating single-line star, a fast rotator and a double-line binary. After the visual inspection, we have omitted 15 stars that have double-line features in their spectra. Eight of them are known spectroscopic binaries: HD~223778, HD~94686, HD~166865, HD~110533, HD~166866, HD~205234, HD~195850, HD~209942 (\citealt{Pourbaix2004, Nordstrom2004}).
For other stars Gaussian fits were made in order to determine a minimum of the profile, and hence the radial velocity as it is shown in Figure~\ref{fig:ccf_gauss_fit}. 
Figure~\ref{fig:vrad_histo} is a histogram of the radial velocities calculated for the 213 stars. The majority of the spectra have radial velocities between $-40$~and~+20~km\,s$^{-1}$. 

\begin{figure}[!htb]
   \advance\leftskip 0cm
      \centering
   \includegraphics[scale=0.26]{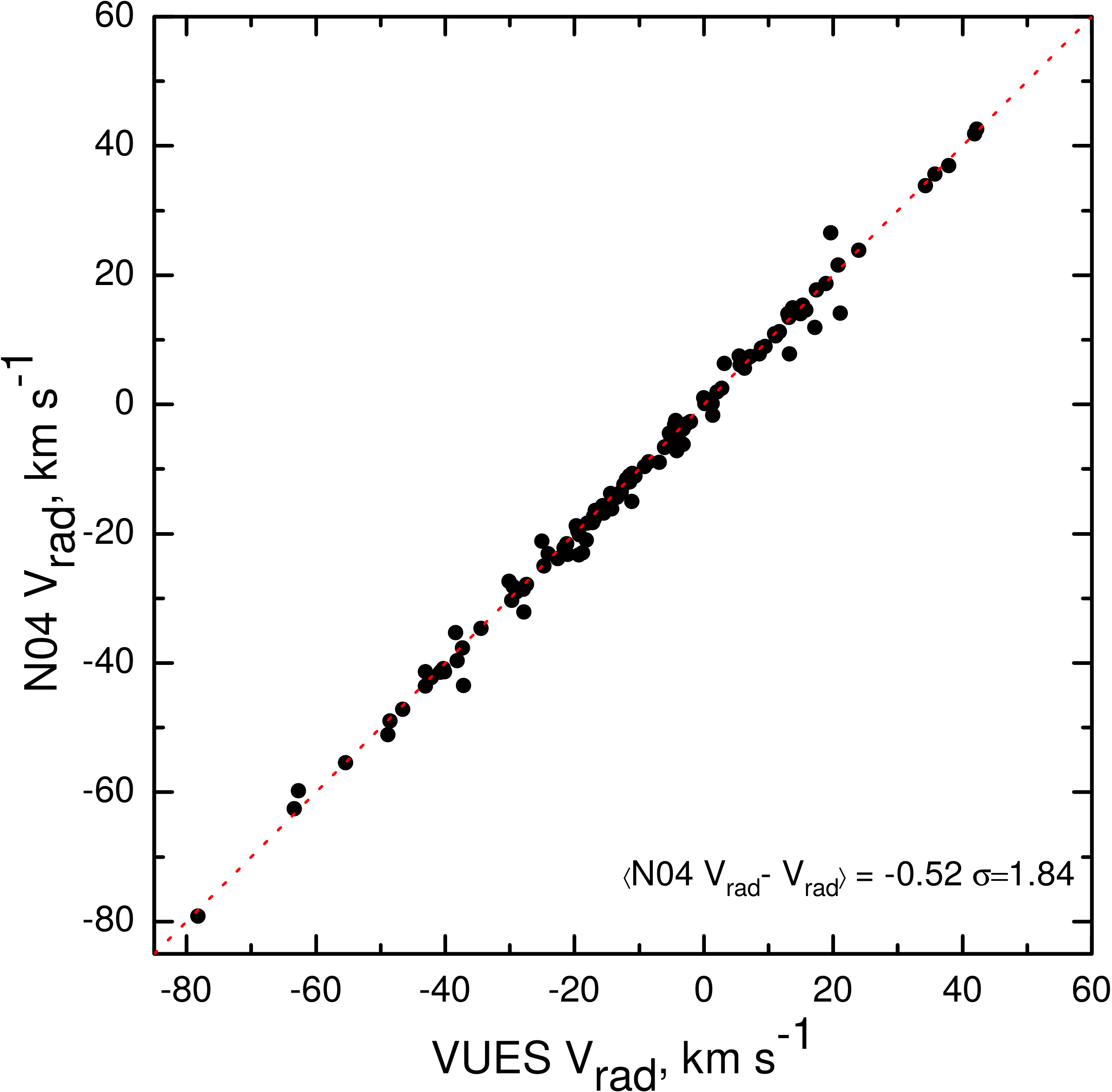}
  \caption{Comparison of radial velocities derived in this study and by \citet{Nordstrom2004} (139 stars). The red dashed line with a slope of 1 is shown for comparison.}
  \label{fig:vrad_nordstrom}
  \end{figure}

\begin{figure}[!htb]
   \advance\leftskip 0cm
      \centering
   \includegraphics[scale=0.26]{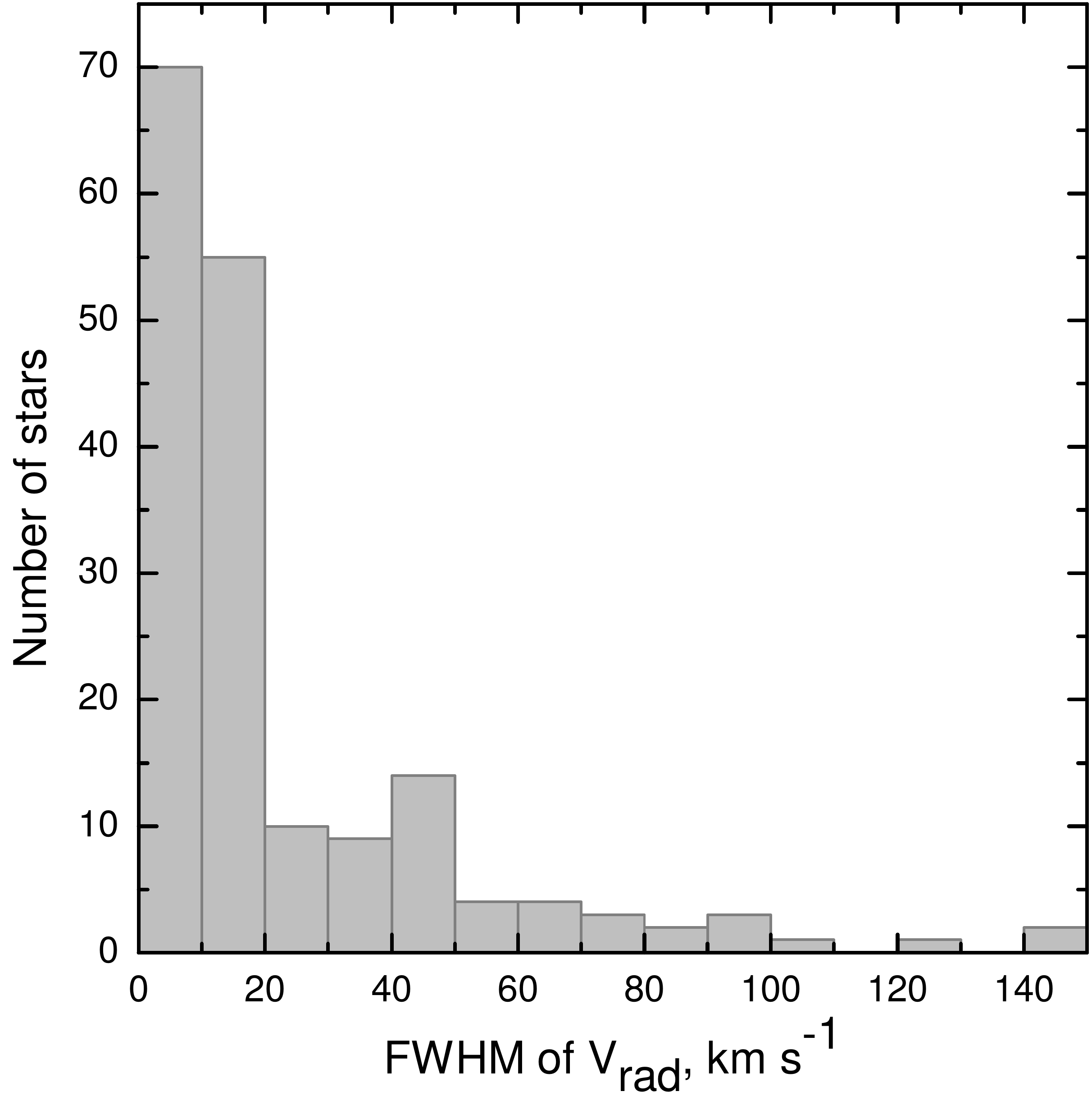}
  \caption{Histogram of the FWHM of the $V_{\rm rad}$ values calculated for all targeted stars.}
  \label{fig:ccfFWHM_histo}
  \end{figure}

Within our dataset, we found 139 stars in common with the \citet{Nordstrom2004} study. Their radial velocity data were obtained with the photoelectric cross correlation spectrometers CORAVEL (\citealt{Baranne1979,Mayor1985}). In Figure~\ref{fig:vrad_nordstrom}, we show a  comparison of the \citet{Nordstrom2004} radial velocity values with those determined in our study. The mean and standard deviation of  differences between the two sets is $\langle \Delta V_{rad} \rangle=0.48\pm1.5$~km\ s$^{-1}$.
Figure~\ref{fig:ccfFWHM_histo} displays a histogram of the FWHM of the CCF (the red dotted line in Figure~\ref{fig:ccf_gauss_fit}). It shows that the majority of spectra returned a FWHM of less than 20~km\ s$^{-1}$. 
This FWHM indirectly reflects the rotational velocity of a star that broadens the observed spectral features depending on the value of the rotational velocity. It was not possible to measure equivalent widths of lines for 51 stars with a satisfactory quality because of the broad and blended lines. All these fast-rotating stars show the FWHM of the CCF larger than 25~km\,s$^{-1}$. \citet{Nordstrom2004} have derived $v$sin$i$ for 28 of our \starcountrotate~fast-rotating stars and for 96 of our \starcountsample~slow-rotating stars. Based on that catalog, 28 of our fast-rotating stars rotate from 19 to 70~km\,s$^{-1}$, while other 96 stars rotate up to 15~km\,s$^{-1}$. 

Therefore we have excluded \starcountbinary~stars that have double-line features in their spectra and \starcountrotate~fast-rotating stars. A comprehensive analysis of these stars will be carried out in a separate work. The final sample consists of \starcountsample~slow-rotating stars, and the atmospheric parameters are provided in this paper for all of them.

\subsection{Kinematic properties}

Kinematic values for our stars were calculated using the python based package for galactic-dynamics calculations \textit{Galpy} \footnote{http://github.com/jobovy/galpy} by \citet{Bovy15}. A sample of kinematic parameters is presented in Table~\ref{tab:kinematics}.  Paralaxes, proper motions and coordinates required for the space velocities and galactic position calculations were taken from the TGAS (\textit{The Tycho-Gaia Astrometric Solution}) catalogue \citep{Michalik15}. The radial velocities of programme stars were determined in this work. To relate the calculated parameters to the Sun, we have used the solar distance from the Galactic plane $z_{\odot}=0.02$~kpc \citep{Joshi07}. The Sun's motion components relative to the local standart of rest ($U_{\odot}$, $V_{\odot}$, $W_{\odot}$) = (11.10, 12.24, 7.25)~km\,s$^{-1}$ were adopted from \cite{Schonrich10}. 
The orbital parameters were calculated using the default potential (\textit{MWPotential2014}) consisting of a bulge, disk and NFW halo (see \citealt{Bovy15}) and the integration time of 5~Gyrs. The rotational velocity of the disk at the Sun's Galactocentric radius in \textit{MWPotential2014} is fixed to $V_{\rm c}=$220~km\,s$^{-1}$ that is adopted from the APOGEE spectroscopic survey data by \citealt{Bovy2012} ($V_{\rm c}=$218$\pm$6~km\,s$^{-1}$). The value $V_{\rm c}=$220~km\,s$^{-1}$ is also recommended by the International Astronomical Union (IAU)  (\citealt{Kerr1986}). However, there are still debates whether it needs to be revised upward to around $V_{\rm c}=$236~km\,s$^{-1}$ (\citealt{Kawata2018}), to $V_{\rm c}=$240~km\,s$^{-1}$ (\citealt{Reid2016}), or even up to $V_{\rm c}=$250~km\,s$^{-1}$ and more (\citealt{Reid2009, Schonrich2012}).

\subsection{Determination of atmospheric parameters}

We used a pipeline of analysis that was constantly used for the Gaia-ESO survey computations by the Vilnius node. It is described in \citet{Smiljanic2014}, here we note the most important information.

\begin{figure}[htb]
   \advance\leftskip 0cm
   \centering
   \includegraphics[scale=0.26]{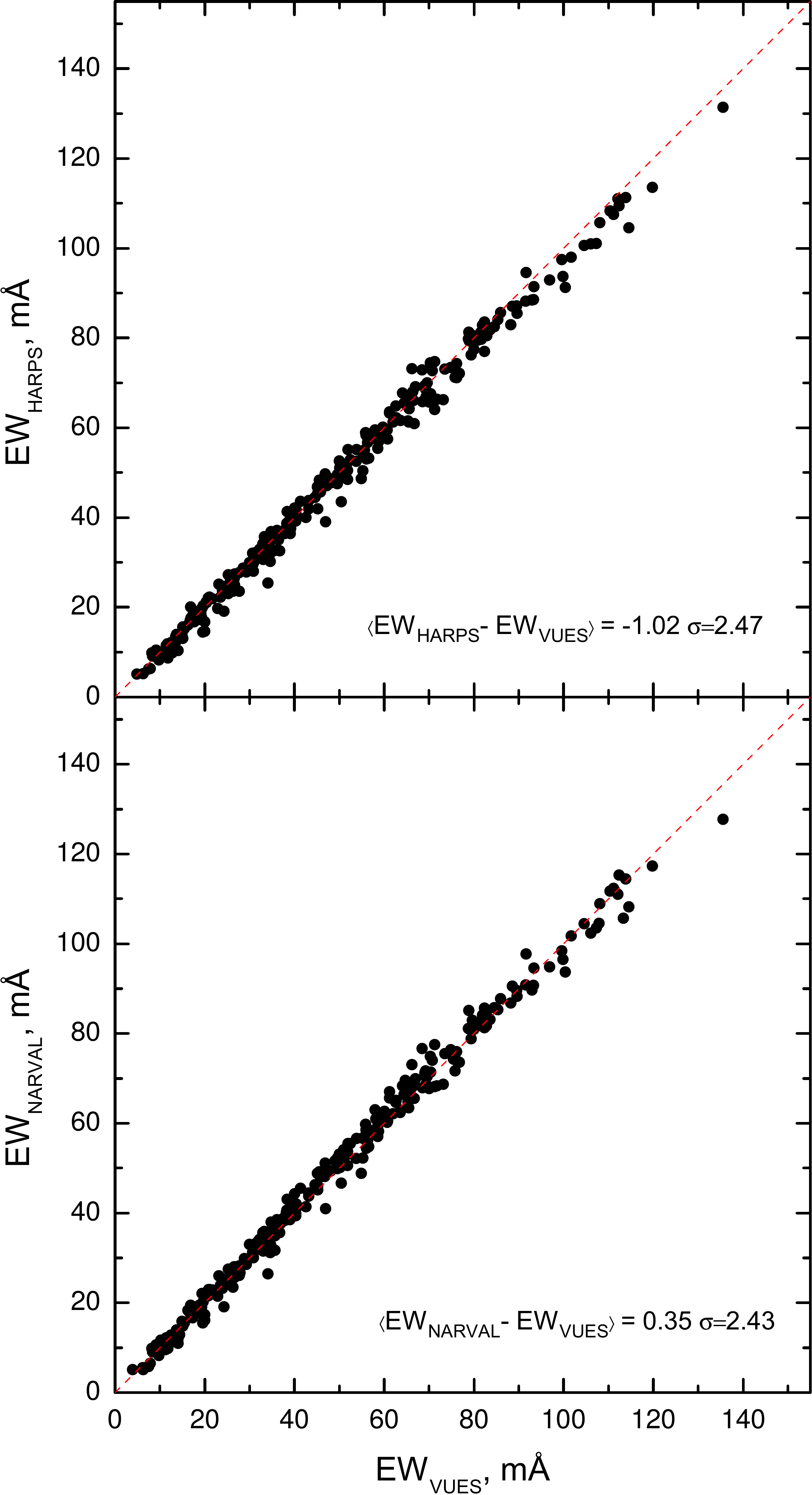}
  \caption{Comparison of equivalent widths of solar iron lines measured in the HARPS, NARVAL and VUES spectra. The red dashed lines with a slope of 1 are shown for comparison.}
  \label{fig:ew_narval_palyginimas}
\end{figure}  

pa
Stellar atmospheric parameters were determined using traditional equivalent width (EW) based methods. EWs were measured using the DAOSPEC (\citealt{Stetson2008}) software.
Figure~\ref{fig:ew_narval_palyginimas} show a comparison of EWs measured for the solar Fe~I lines  from the VUES spectra and the HARPS and NARVAL spectra, respectively. The EWs measured from spectra of these spectrographs agree very well. 

Effective temperatures were determined by minimizing a slope of abundances obtained from Fe~I lines with respect to the excitation potential. Surface gravities were determined by forcing the measured Fe~I and Fe~II lines to yield the same iron abundance. Microturbulent velocities were determined by forcing Fe~I abundances to be independent of the EWs of the lines.  A custom wrapper software was developed to measure EWs, and compute the main atmospheric parameters and abundances automatically. 
  
The Fe~I and Fe~II linelist (299 lines) was taken from \citet{Sousa2008} with . The stellar atmospheric parameters were computed using the 10-th version of the MOOG code (\citealt{Sneden1973}) using a grid of MARCS stellar atmosphere models (\citealt{Gustafsson2008}). The final interpolated model in the WEBMARCS format for MOOG was calculated using a modified interpolation software, provided together with the MARCS models.

The pipeline performs an iterative sequence of abundance calculations in order to minimize the abundance dependency on the line excitation potential, [Fe~I/Fe~II] and $\sigma$([Fe~I/H]). It is done by  employing the Nelder-Mead method (\citealt{Nelder1956}).
Every resulting abundance for every single line that departed from the mean value by more than 2\,$\sigma$ was flagged as an outlier.

\subsection {Error estimation for the atmospheric parameters}
\label{sec:errorsatmospheres}

Using these previously described techniques, we also estimated the different sources
of possible uncertainties raising in deriving stellar atmospheric parameters. 
Typically there are multiple sources of uncertainties, some of which affect
single lines independently (e.g. random errors of the line fitting or continuum
placement) as well as errors of the employed method, which are mainly associated with the uncertain linear regression fit or the atomic parameters.

First, we studied the errors that might be caused by a possible incorrect
continuum placement. There are several ways to evaluate these errors. One 
is to perform the Monte Carlo simulations, selecting a statistically significant set of spectra and adding a noise artificially. This simulation can show the sensitivity of the method to noises and continuum placement. Another way is to follow the line-to-line scatter.

If there is a statistically significant number of lines for a given element, the scatter informs
about the combined effect of the erroneous continuum
placements, equivalent width measurements, and uncertain atomic parameters for different lines.

For the Monte Carlo simulations, we took spectra of two stars that are most typical for the sample: TYC~4573-1916-1 ($T_{\rm eff}$=6153~K, log\,$g$=4.01, [Fe/H]=$-0.07$) and the Sun ($T_{\rm eff}$=5779~K, log\,$g$=4.49, [Fe/H]=$-0.03$)\footnote{We note that the Solar values of our method ($T_{\rm eff}$=5779$\pm24$~K, log\,$g$=4.49$\pm0.12$, [Fe/H]=$-0.03\pm$0.04) were computed while analysing the Solar spectrum of VUES the same way as any other star in the programme. 
The values are very close to the nominal Solar values declared in 2015 by the IAU resolution B3 ($T_{\rm eff}$=5772.0~K, log\,$g~=~4.438068$, \citealt{Prsa2016}) or the Gaia benchmark star solution ($T_{\rm eff}$=5771~K, log\,$g$=4.4380, \citealt{Heiter2015b}). Both \citet{Heiter2015b} and \citet{Prsa2016} studies benefit from very precise methods that are independent of spectroscopy and atmospheric models and reach errors in $T_{\rm eff}$ and log\,$g$ up to 1~K and 0.0002~cm~s$^{-1}$, respectively. Similarly, the metallicity value could be slightly off 0.0~dex in different methods (e.g. Table~2~in~\citealt{Jofre2014}).}.
The spectra of these bright stars have a signal-to-noise ratio (SNR) of more than 200 per pixel. In order to investigate a possible noise influence, we degraded these spectra with a white Gaussian noise to the SNR equal to 25, 50, and 75 per pixel. We then generated 100 spectra for each SNR value, remeasured equivalent widths and derived the corresponding 100 atmospheric parameters for each SNR. In that way, we determined sensitivity of our atmospheric parameters to the SNR. These uncertainties are provided in a form of the standard deviations in Table~\ref{tab:montecarlo}.

\begin{table}
\caption{Errors due to the uncertain continuum placement and equivalent width measurement. Based on the Monte Carlo simulations.}
\centering
\begin{tabular}{lrrr}
\hline\hline
 & S/N=25 & S/N=50 & S/N=75 \\
\hline
\multicolumn{4}{c}{TYC~4573-1916-1}\\
\multicolumn{4}{c}{$T_{\rm eff}=6153$\,K, ${\rm log}~g=4.01$, ${\rm [Fe/H]}=-0.07$}\\

$\sigma_{T_{\rm eff}}$	&	37	&	31	&	28	\\
$\sigma_{{\rm log}~g}$ 	&	0.08	&	0.08	&	0.08	\\
$\sigma_{{\rm [Fe/H]}}$  &	0.02	&	0.02	&	0.01	\\
$\sigma_{v_{\rm t}}$  	&	0.08	&	0.07	&	0.04	\\

\hline
\multicolumn{4}{c}{Sun$^*$}\\
\multicolumn{4}{c}{$T_{\rm eff}=5779$\,K, ${\rm log}~g=4.49$, ${\rm [Fe/H]}=-0.03$}\\
$\sigma_{T_{\rm eff}}$	&	52	&	36	&	35	\\
$\sigma_{{\rm log}~g}$ 	&	0.10	&	0.08	&	0.07	\\
$\sigma_{{\rm [Fe/H]}}$  &	0.06	&	0.01	&	0.01	\\
$\sigma_{v_{\rm t}}$  	&	0.10	&	0.08	&	0.06	\\

\hline
 \label{tab:montecarlo}
\end{tabular}

$^*$Solar atmospheric parameters derived with our method (see Section~\ref{sec:errorsatmospheres}). 
\end{table}  

  \begin{figure}[!htb]
   \advance\leftskip 0cm
      \centering
   \includegraphics[scale=0.30]{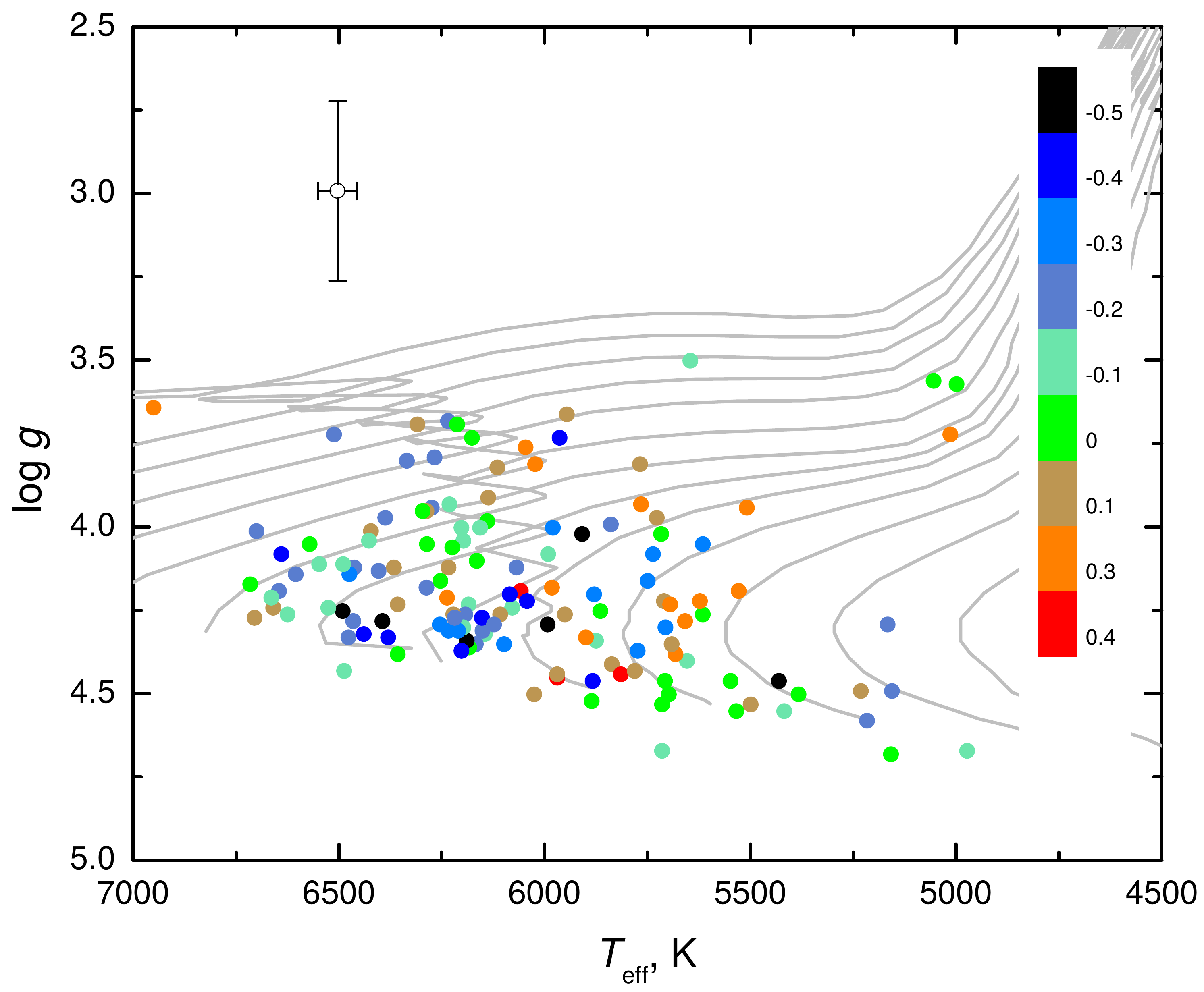}
  \caption{Temperature--gravity diagram of investigated stars (dots) with metallicity coded in color. Evolutionary tracks by \citet{Girardi2000} with masses between 0.7 and 1.9~$M_{\odot}$ and $Z_{\rm ini}$=0.019 are plotted by the grey solid lines. }
  \label{fig:hr}
  \end{figure}

However, this kind of error estimation is only robust when there are many lines. We were able to use 92--252 Fe~I and 7--32 Fe~II lines in the analysis. The line-to-line scatter of Fe~I and Fe~II abundances was propagated to the uncertainties of atmospheric parameters as follows:
\begin{itemize}
 \item The uncertainty for the effective temperatures was estimated by obtaining the boundary temperature values of the possible satisfactory parameter space, using
the error of the linear regression fit to the Fe~I abundances.
 \item The uncertainty of the surface gravity was obtained searching for the boundary values of log\,$g$ according to the standard deviations of abundances from the Fe~I and Fe~II lines.
 \item The uncertainty of the microturbulent velocity was obtained by employing the error of the standard deviation of the neutral iron abundances.
 \item The [Fe/H] standard deviation ($\sigma$[Fe~I/H]) was adopted as the metallicity uncertainty.
\end{itemize}
These uncertainties are provided for every star in Table~\ref{tab:results} as well.

\begin{table*}
\caption{Sample table of the atmospehric parameters of the programme stars.}

\begin{tabular}{lccccccccccccccccc}
\hline\hline

Tycho2 ID	&	$T_{\rm eff}$	&	$\sigma_{T_{\rm eff}}$	&	${\rm log}\,g$	&	$\sigma_{{\rm log}\,g}$	&	$\rm{[Fe/H]}$	&	$\sigma_{{\rm [Fe/H]}}$	&	$v_{\rm t}$	&	$\sigma_{v_{\rm t}}$	\\

 &  K	&	K	&		&		&		&		&	km\ s$^{-1}$	&	km\ s$^{-1}$		\\
	\hline
TYC 4141-1496-1 &	6353	&	54	&	4.24	&	0.35	&	0.01	&	0.08	&	1.39	&	0.24	\\
TYC 4141-589-1 	&	6711	&	93	&	4.18	&	0.36	&	$-0.10$	&	0.12	&	1.76	&	0.34 \\
TYC 4160-145-1 &	6042	&	31	&	3.77	&	0.22	&	0.10	&	0.08	&	1.33	&	0.24\\
\hline
 \label{tab:results}
\end{tabular}

Notes. This is a small part of the table (the full table is available at the CDS).\\
\end{table*}

\section{Stellar parameters}
\label{sec:parameters}

\subsection{Atmospheric parameters}
\label{sec:atmosphericparameters}

\begin{figure*}[!htb]
   \advance\leftskip 0cm
      \centering
   \includegraphics[scale=0.26]{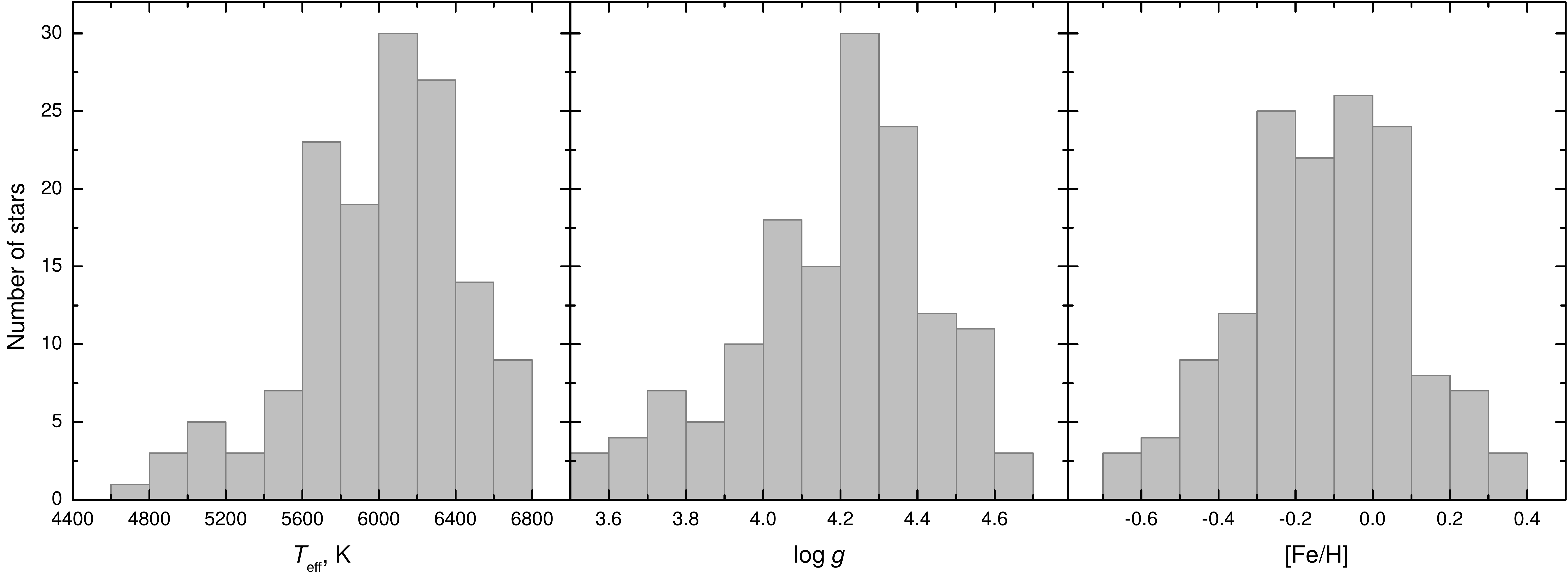}
  \caption{Histograms of the determined spectroscopic parameters ($T_{\rm eff}$, log\,$g$,  and [Fe/H]) for all stars in our sample.}
  \label{fig:resultsHISTOGRAM}
  \end{figure*}

We have determined values of effective temperature, surface gravity, metallicity, microturbulent velocity, and radial velocity with their corresponding uncertainties for \starcountsample~stars and the sample of results is listed in Table~\ref{tab:results}. The final stellar parameters are shown in the ($T_{\rm eff}$, ${\rm log}\,g$)-diagram with color-coded metallicity (Figure~\ref{fig:hr}). Also we plotted stellar evolutionary tracks in the background (\citealt{Girardi2000}) with masses between 0.7 and 1.9~$M_{\odot}$ and the initial metallicity $Z_{\rm ini}$=0.019. Majority of the investigated stars have masses between 0.9 and 1.5~$M_{\odot}$.   
Distributions of the determined $T_{\rm eff}$, ${\rm log}\,g$, and [Fe/H] are shown in Figure~\ref{fig:resultsHISTOGRAM}. 

The determined $T_{\rm eff}$ span from 4700~K to 6950~K (Figure~\ref{fig:resultsHISTOGRAM}a) with a peak at 6100~K, ${\rm log}\,g$ range 
from 3.5 to 4.7 (Figure~\ref{fig:resultsHISTOGRAM}b) peaking at 4.3, and [Fe/H] are from $-0.7$ to +0.4~dex (Figure~\ref{fig:resultsHISTOGRAM}) with majority of stars being of solar metallicity.  

Several studies in the literature have derived stellar parameters for some stars of our sample. Hence, we collected those results and compared them with the ones obtained in our study. In the $fe\_h$ catalogue of the SIMBAD database and the PASTEL catalog (\citealt{Soubiran2010}) we found spectroscopic parameters for 39 common stars (27\% of our sample). 
It was not possible to find a significant number of atmospheric parameters from a single uniform spectroscopic study. Thus we selected the $T_{\rm eff}$, ${\rm log}~g$, and [Fe/H] spectroscopic determinations which are relatively recent: \citet{Galeev2004, Delgado2015, Mishenina2012, Mishenina2013, Guillout2009, Gonzalez2010, Takeda2007a, Takeda2007b, Prugniel2011, Ramirez2013, Fuhrmann2008, Feltzing1998, Fuhrmann2004, Chen2000, Gray2003, Valenti2005}. This sample of 39 stars we call the spectroscopic comparison sample.

\begin{figure*}[htb]
 \advance\leftskip 0cm
 \centering
 \includegraphics[scale=0.22]{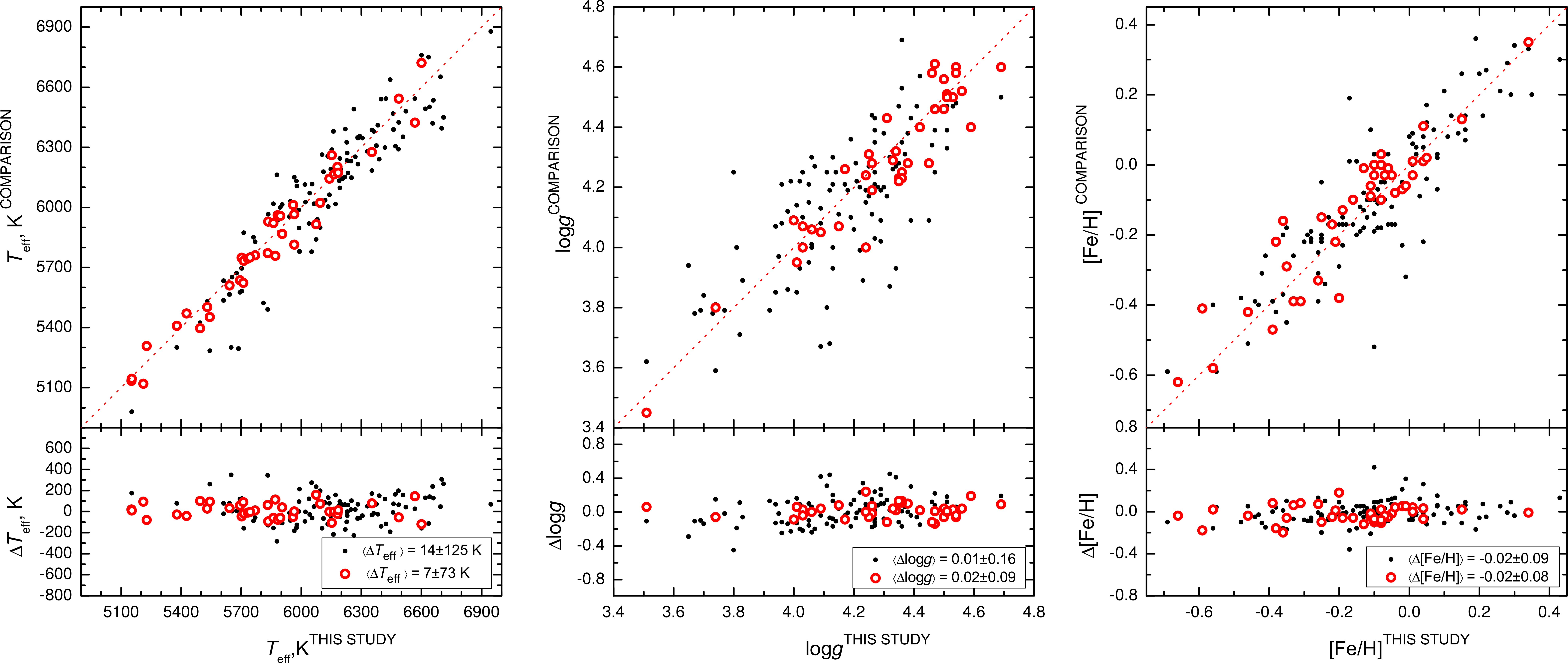}
  \caption{Comparison between $T_{\rm eff}$, ${\rm log}~g$, and [Fe/H] values of our study with values from the spectroscopic comparison sample (39 stars, red circles) and with values from the photometric comparison sample (119 stars, black dots). The red dashed lines with a slope of 1 are shown for comparison. See the text for more information.}
  \label{fig:resultsCOMAPRISON}
\end{figure*}

Figure~\ref{fig:resultsCOMAPRISON} shows a comparison between $T_{\rm eff}$, ${\rm log}~g$, [Fe/H] values of our sample and the spectroscopic comparison sample of 39 stars (red circles). Another so called photometric comparison sample of 119 stars (black dots) was selected from the study by \citet{Casagrande2011}. The effective temperatures in this work are based on the infrared flux method and should be quite accurate.  

Effective temperature consistency between our sample and the spectroscopic comparison sample of 39 stars is quite good ($\langle \Delta T_{\rm eff} \rangle$=14$\pm$73~K).
The photometric comparison sample is consistent with our results as well ($\langle \Delta T_{\rm eff} \rangle=14\pm$125~K). The identical  mean difference between photometric and spectroscopic determinations, $T_{\rm eff}= 13\pm 95$~K, was found by \citet{Casagrande2011} using their sample of 1522 stars. 

Surface gravities are consistent with both photometric ($\langle \Delta {\rm log}~g \rangle=0.01\pm$0.16) and spectroscopic ($\langle \Delta {\rm log}~g \rangle=0.02\pm0.09$) comparison samples.

Metallicities show the same negligible bias ($\langle \Delta {\rm [Fe/H]} \rangle=-0.02\pm0.09$~dex) for both spectroscopic and photometric comparison samples. The metallicity distribution in our sample of stars meets the one typically observed in the solar neigborhood. For example, \citet{Prieto2004} found $\langle{\rm [Fe/H]}\rangle=-0.11\pm0.18$ which is identical to our value  $\langle{\rm [Fe/H]}\rangle=-0.11\pm0.20$.

  \begin{figure*}[!htb]
   \advance\leftskip 0cm
      \centering
   \includegraphics[scale=0.60]{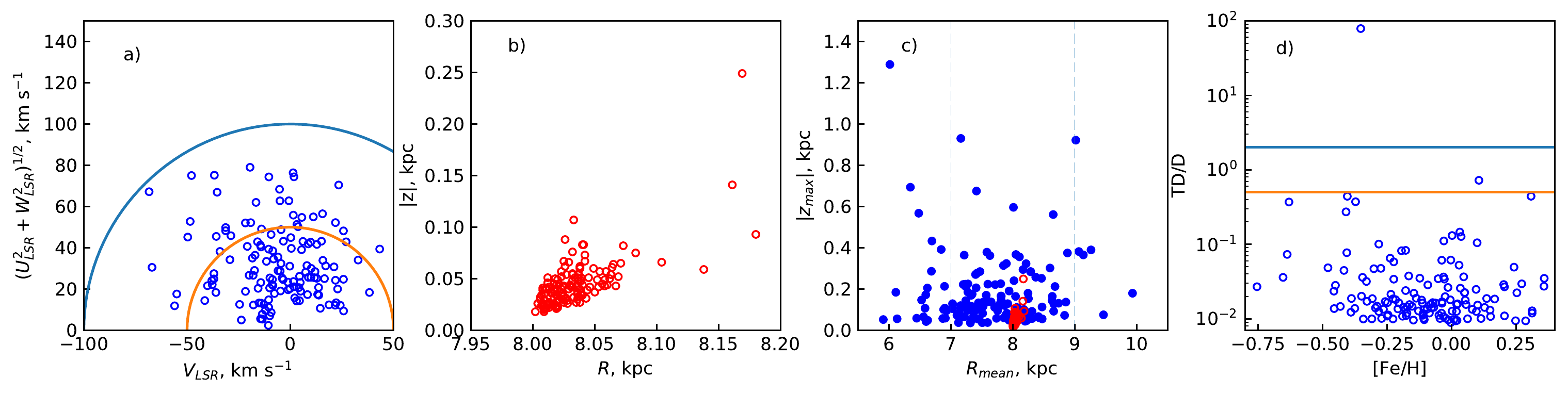} 
  \caption{Kinematic parameters of our sample stars. a) Toomre diagram of our sample stars, lines show constant values of the total space velocity ($v_{\rm tot}=(U_{\rm LSR}+V_{\rm LSR}+W_{\rm LSR})^{\rm 1/2)}$) at 50 and 100 km\,s$^{\rm -1}$, b) distribution of sample stars in the $|z|$ plane vs. $R_{\rm gc}$, c) distribution of sample stars in $z_{\rm max}$ vs. $R_{\rm mean}$ plane where two vertical dashed lines delimit the solar neighborhood 7$<R_{\rm gc}<$9~kpc and red circles are the ones from plane b for comparison, d) kinematical thick disk-to-thin disk probability ratios (TD/D) versus metallicity for the sample stars upper and lower lines mark TD/D=2.0 and TD/D=0.5, respectively. 
}
\label{fig:resultsKINEMATICS}
  \end{figure*}  
  
\subsection{Kinematic parameters}

In Figure~\ref{fig:resultsKINEMATICS} we present the kinematical distribution of sample stars. According to the distribution of stars in the Toomre diagramme (the plane $a$ in  Figure~\ref{fig:resultsKINEMATICS}), our sample stars are well confined inside $v_{\rm tot}=100$~km~s$^{\rm -1}$, however, most of them are clearly inside $v_{\rm tot}=50$~km~s$^{\rm -1}$ ($v_{\rm tot}=(U_{\rm LSR}+V_{\rm LSR}+W_{\rm LSR})^{\rm 1/2)}$). Currently all these stars are well confined close (distances up to 0.366~kpc) to the Sun (see plane $b$ in Figure~\ref{fig:resultsKINEMATICS}). In the $R_{\rm mean}$ vs. $z_{\rm max}$ plane our sample stars have mean galactocentric distances from 6 to 10~kpc and vertical distances up to 1.3~kpc (the plane $c$ in Figure~\ref{fig:resultsKINEMATICS}). The plane $d$ in  Figure~\ref{fig:resultsKINEMATICS} shows the thick-to-thin disk probability ratios (TD/D) versus metallicity, computed the same way as in \citet{Bensby2003, Bensby2014}, who show that stars with TD/D~$>$~2 are potential thick disk stars, stars with TD/D~$<$~0.5 potentially belong to the thin disk, and stars with 0.5~$<$~TD/D~$<$~2.0 are called "in-between stars".

There are 28 stars in our sample that are kinematically hot visitor stars from the inner ($R_{\rm mean}<$7~kpc, 22 stars) and outer ($R_{\rm mean}>$9~kpc, 6 stars) Galactic regions that are passing through the Solar neighborhood (7$<R_{\rm gc}<$9~kpc, \citealt{Carigi2015, Rojas2016}) on highly eccentric orbits (see plane $c$ in Figure~\ref{fig:resultsKINEMATICS}). The velocity dispersions of the sample stars are $\sigma_{\rm U}=32$~km\,s$^{\rm -1}$, $\sigma_{\rm V}=21$~km\,s$^{\rm -1}$ and $\sigma_{\rm W}=14$~km\,s$^{\rm -1}$, that are very close to the characteristic velocity dispersions (see \citealt{Bensby2014}) of the thin disk. All of our sample stars are confined inside the $v_{\rm tot}=100$~km\,s$^{\rm -1}$ according to the Toomre diagramme, and only two of our sample stars have TD/D higher than 0.5, thus we can expect that our sample of stars is a characteristic thin disk sample. The contamination from the thick disk should be very small. 

\begin{figure*}[!htb]
 \advance\leftskip 0cm
 \centering
 \includegraphics[scale=0.80]{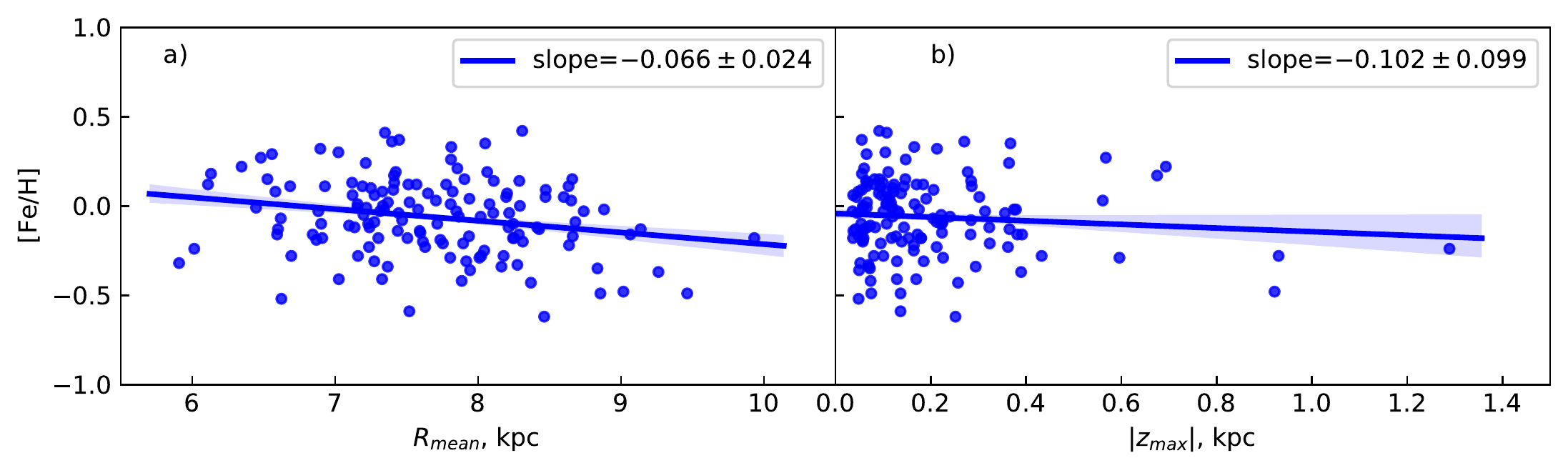} 
 \caption{Metalicities of the sample stars as a function of the mean galactocentric radii and b) vertical distribution. The solid lines show linear fits to the data while shaded areas marks 1$\sigma$ uncertainties of the fits.
}
\label{fig:resultsGRADIENTS}
\end{figure*}

\subsection{Spatial metallicity distribution of sample stars}

In Figure~\ref{fig:resultsGRADIENTS} we present a metallicty distribution in the Galactic disk according to $R_{\rm mean}$  and $|z_{\rm max}|$ where the solid line is a fit to the data with the $\pm$1$\sigma$ confidence bands shown as the shaded ares. A slope of the fit of the radial distribution is $-0.066\pm0.024$~dex~kpc$^{-1}$, thus the radial metallicity gradient of our data is significant by about 2.5\,$\sigma$. The vertical distribution is $-0.102\pm0.099$~dex~kpc$^{-1}$, thus the vertical metallicity gradient is slightly negative, but significance of the slope is low, only 1\,$\sigma$.

The radial chemical gradients of the thin disk have been studied in a number of papers during the recent years that have used different selection criteria to purify a thin disk population from thick disk memebers. The RAVE survey (\citealt{Coskunoglu2012}, \citealt{Bilir2012}) provided metallicity gradients separately for giant and dwarf stars for most probable thin-disk stars and found $-0.033\pm0.007$~dex~kpc$^{-1}$. \citet{Boeche2013} found the metallicity gradient to be $-0.065\pm0.003$~dex~kpc$^{-1}$ for the most probable thin disk sample. A very similar result was also found in the SEGUE survey (\citealt{Cheng2012}), a slope of $-0.066\pm0.037$~dex~kpc$^{-1}$ for their sample which was dominated by the thin-disk stars. The Gaia-ESO survey delivered similar gradients in a number of papers: $-0.045\pm0.009$~dex~kpc$^{-1}$ by \citet{Mikolaitis2014}, $-0.058\pm0.008$~dex~kpc$^{-1}$ by \citet{Blanco2014} and $-0.068\pm0.016$~dex~kpc$^{-1}$ by \citet{Bergemann2014}.  \citet{Anders2014} in their study of the APOGEE survey found radial gradients for their HQ and Gold samples to be $-0.066\pm0.006$~dex~kpc$^{-1}$ and $-0.074\pm0.01$~dex~kpc$^{-1}$, respectively. The same authors repeat similar gradients for samples dominated by thin disk stars in \citet{Anders2017}. Also results from cepheids should be mentioned. Cepheid variables are younger than 200~Myr and are clearly thin disk stars. Their radial gradients span from $-0.029$~dex~kpc$^{-1}$ (\citealt{Andrievsky2002}) to $-0.052\pm0.003$~dex~kpc$^{-1}$ (\citealt{Lemasle2008, Pedicelli2009}).

From all this information we see that differences in metallicity galactocentric gradients can be significant. The reason could be in strategy of building a sample. The results of our sample mostly agree within errors with \citet{Bergemann2014}, \citet{Cheng2012}, \citet{Boeche2013}, and \citet{Anders2014}. However, we note that all above-mentioned studies, except \citet{Anders2014}, computed the gradients according to a current stellar position. \citet{Anders2014} gradients are computed according to a median Galactocentric distance, specifically attributing their study to stars that compose the current Solar neighborhood, thus their results should be more clearly comparable to ours. In general, the chemical gradients of the Galactic disk with respect to $R_{\rm mean}$ or $|z_{\rm max}|$ have been rarely studied before.

The vertical metallicity gradient derived by \citet{Bilir2012} in the thin disk is $-0.109\pm0.008$~dex~kpc$^{-1}$. \citet{Boeche2013} observed a slow metallicity decrease in the range
$0.0<|z_{\rm max}|<1.0$, where the thin disk is dominant. According to the results by \citet{Duong2018}, the thin disk exhibits a steep negative vertical metallicity gradient, $-0.18\pm0.01$~dex~kpc$^{-1}$. Our results are in broad agreement with the abovementioned studies, however, we note, that gradients of \citet{Bilir2012} and \citet{Duong2018} are computed according to current stellar positions while ours is according to $z_{\rm max}$.

The formation of the Galactic disk is beyond the main scope of this study. But we must mention that even if we study a small sample of stars that are located to quite a narrow direction of the Galaxy, we still can sense the Galactic disk formation history. The observed radial and vertical stellar abundance distributions are results of the Galactic disk formation processes and can be compared with thin disk evolutionary models. Our elemental abundance radial gradients agree with the models of \citet{Cescutti2007}, which assumed an inside-out build-up of the disk on a time-scale of 7~Gyr in the solar neighborhood (see \citet{Matteucci1989, Cescutti2007, Pilkington2012}). On the other hand, it is interesting to note that \citet{Haywood2013} advocated an outside-in formation of the disk that also should create the negative metallicity gradients in the thin disk. Our results broadly agree with the model by \citet{Toyouchi2018}, who studied the gas infall, re-accretion of out-flowing gas, and radial migration of disk stars, as well as with the thin-disk chemical evolution models by \citet{Minchev2013, Minchev2014}, which modelled an  inside-out disk formation including stellar migration, triggered by mergers in the early epochs and afterwards by secular processes of the Galactic bar and spiral arms.

\section{Summary}
\label{sec:summary}

This paper is the first data release of the Spectroscopic and Photometric Survey of the Northern Sky (SPFOT) that aims to provide a detailed chemical composition from high-resolution spectra and photometric variability data for bright stars in the northern sky using telescopes of the Mol\.{e}tai Astronomical Observatory, Vilnius University. 

We have observed high-resolution spectra for all 213 photometrically selected 6500~K and cooler dwarfs with magnitudes up to $V=8$~mag in the field with radius of 20~degrees towards the northern celestial pole.
This region of the sky is very important since it will be intensively studied by the NASA~TESS mission (\citealt{Ricker2015, Sullivan2015}) which is an  important test-bench for the PLATO~2.0 space mission. 

We determined spectroscopic atmospheric parameters for 140 slowly rotating stars (for 73\% of the stars spectroscopic parameters were determined for the first time). Our results have no systematic differences when compared  with other recent studies. Other 73 stars, of the sample of 213, were either fast rotators or double-line binaries and will be analyzed later using different techniques. 

We also determined a number of kinematic parameters that confirmed that almost all of the investigated stars belong to the kinematic thin disk. We employed the mean radial and maximal vertical distances and confirmed that metallicity spatial distributions of bright dwarfs located towards the northern celestial pole comply with the latest inside-out thin disk formation models, including those with migration taken into account. 

Having in mind that only 30\% of bright ($V<8$~mag) FGK main sequence stars have spectroscopic determinations of atmospheric parameters, we emphasize the importance of high-resolution spectral analysis of bright stars, especially for the 'hot'-regions of the sky such as the TESS continuous viewing zones (\citealt{Sharma2017}), $Kepler$ and $K2$ fields (\citealt{Molenda2017, Niemczura2017, Petigura2017}), and the PLATO fields (\citealt{Miglio2017}).

\begin{acknowledgements}

We thank the anonymous referee for comments and suggestions, which helped to improve this Paper significantly.
This research has made use of the SIMBAD database and NASA’s Astrophysics Data System (operated at CDS, Strasbourg, France), and
was funded by the grant from the Research Council of Lithuania (LAT-08/2016).
\end{acknowledgements}



\end{document}